\documentstyle[prl,aps,epsfig,multicol]{revtex}
\begin{document}

\def\be{\begin{equation}}
\def\ee{\end{equation}}
\def\bearr{\begin{eqnarray}}
\def\eearr{\end{eqnarray}}
\def\tc{$T_c$}
\def\tcl{$T_c^{1*}$}
\def\c2{$ CuO_2$}
\def\lsco{LSCO}
\def\bi{Bi-2201}
\def\tl{Tl-2201}
\def\hg{Hg-1201}   

\title{ The Principle of Valence Bond Amplitude Maximization in
Cuprates:\\
How it breeds Superconductivity, Spin and Charge Orders}
\author{G. Baskaran}

\address{The Institute of Mathematical Sciences\\
C I T Campus\\
Madras 600 113, India}
\maketitle
\begin{abstract}
A simple microscopic principle of `Valence bond (nearest neighbor 
singlet) amplitude maximization '(VBAM) is shown to be present 
in undoped and optimally doped cuprates and unify the very different 
orderings such as antiferromagnetism in the Mott insulator 
and the robust superconductivity accompanied by an enhanced 
charge and stripe correlations in the optimally doped 
cuprates. VBAM is nearly synonymous with the energy minimization 
principle. It is implicit in the RVB theory and thereby makes 
the predictions of RVB mean field theory of superconductivity
qualitatively correct. 
\end{abstract}

\begin{multicols}{2}[]

The qualitative predictions of the RVB mean field theory of high Tc 
superconductivity\cite{bednorz} in cuprates,including the symmetry 
of the order parameter\cite{pwascience,pwabook,bza,kotliar} has turned 
out to be in good 
agreement with experiments. In spite of its approximate character it has
definitely put us in the correct Hilbert space by focusing on the 
key singlet correlations. On the other 
hand various orders, that at a superficial level appear to be 
unusual from the RVB theory point of view, have emerged both 
experimentally and theoretically.  The aim of the present letter 
is to identify an unique cause behind the 
AFM order in the Mott insulator, the enhanced charge and spin  
stripe correlations and low energy spin fluctuations 
in the optimally doped superconducting cuprates
\cite{zaanen,kivelson,tranquada}.  A simple principle of 
`valence bond (nearest neighbor
singlet) amplitude maximization' (VBAM) is shown to emerge as a fairly 
simple cause.  This maximization respects certain geometrical 
and dynamic constraint provided by our strongly correlated system.  
In the case of the Mott insulator, while super exchange interaction
respects VBAM, an irreducible amount of local triplet fluctuations 
is generated in order to satisfy certain geometric constraints. These 
triplet fluctuations interact and condense at momentum $(\pi,..\pi)$ 
resulting in long range AFM order for any dimension $\geq 2$. 
In the optimally doped Mott insulator the constraints are modified
by the charge dynamics. And VBAM continues as an approximate 
principle and produces new orders such as the robust superconductivity 
and the enhanced quasi static charge and spin stripe correlations.
We also argue that VBAM is responsible for the qualitative success
of the RVB mean field theory.

At a more fundamental level the 2 dimensionality, the one band character, 
strong correlation and closeness to a Mott insulating state are 
believed to be responsible for various interesting low energy 
electronic phenomenon in cuprates\cite{pwabook}. The VBAM  hypothesis 
is an emergent consequence of the above and perhaps an useful 
guiding principle that at least helps us to separate cause from effects.  
One of the messages of this letter is that it is VBAM, rather than 
the development of charge or spin stripe correlations that is 
providing a mechanism for superconductivity: on the contrary whenever 
stripes are nurtured it is at the expense of 
superconductivity\cite{gbstripe,scalapinostripe}. 

Our VBAM  hypothesis is proved for the case of undoped Mott insulator by 
simply using the principle of energy minimization in the ground state. 
For the optimally doped Mott insulator various arguments are provided 
in support of this principle. The existence of the sharp 41 meV triplet 
resonance\cite{keimer} at $(\pi,\pi)$ at optimally doped bilayer YBCO 
and its counterpart at various dopings all the way up to zero doping 
is presented as an experimental support for our VBAM hypothesis in 
the optimally doped regime.

The phase coherent resonance of singlet bonds is at the heart of 
the RVB theory of superconductivity. Anderson's RVB wave 
function\cite{pwascience}, 
\be
|RVB\rangle = P_G (\sum a_{ij} b^\dagger_{ij})^{
\frac{N_e}{2}} |0\rangle
\ee
by construction has enhanced VB amplitude. It is a remarkably universal
wave function in the sense, it can describe an antiferromagnetically 
ordered Mott insulating state and the robust superconducting state
of the optimally doped Mott insulator with equal ease for appropriate  
choices of the pair function $a_{ij}$. We believe that this universal
feature is a consequence of the VBAM principle. Also without changing
the RVB wave function and consequently the superconducting property in 
a fundamental fashion, the charge stripe and spin stripe order can be
incorporated by a modulation of the pair function $a_{ij}$.   

To begin with we develop a heuristic picture of VBAM.  In a free 
fermi gas, any short range singlet correlation contained in the ground 
state is a consequence of Pauli principle rather than interactions. 
In the large U Hubbard model, accepted as a good model 
for our narrow band conducting cuprates close to half filling, every 
elementary collision between two electrons tries to establish 
a nearest neighbor singlet correlations (a valence bond). That is,
the virtual transitions to doubly occupied state on a given copper site 
lowers the energy (compared to the $ U = \infty$ case) by the super 
exchange energy 
$J \approx {2t^2 \over U}$ and stabilizes the spin singlet state rather 
than  a triplet state. Elementary collisions, in addition to the Pauli 
principle induce singlet correlations.  Close to half filling elementary 
two body collisions are more frequent than free hopping of charges.  
{\em The on site collision induced valence bond proliferation} is 
at the heart of the VBAM principle.

Let us try to understand the development of long range AFM order in
the spin half Mott insulators in the light of the principle of
VBAM. In an isolated pair of neighboring orbitals, the electron pair 
has a spin singlet (non-magnetic) ground state, through the super 
exchange interaction.  An important question is how this local non magnetic 
singlet tendency manifests itself in a 2 or 3 dimensional lattice.
The spin half Heisenberg Hamiltonian with nearest neighbor interaction 
\be
H =  J \sum_{\langle ij \rangle} ({\bf S}_i \cdot {\bf S}_j - {1\over4}),
\ee
in terms of the underlying electron variables $c's$  and the bond
singlet operator $b^\dagger_{ij} \equiv {1\over \sqrt{2}}
(c^\dagger_{i\uparrow} c^\dagger_{j\downarrow}  - c^\dagger_{i\downarrow}
c^\dagger_{j\uparrow})$ takes the form 
\be
H = - J\sum_{\langle ij \rangle } b^\dagger_{ij} b^{}_{ij},~~~~ 
\mbox{ with } ~~n_{i\uparrow} + n_{i\downarrow} \neq 2
\ee
We have used the important identity\cite{bza} 
\be
({\bf S}_i\cdot {\bf S}_j - 
{1\over4} n_i n_j) \equiv b^\dagger_{ij} b^{}_{ij}  
\ee
The singlet number operator $b^\dagger_{ij} b^{}_{ij}$.  
has an eigen value of 0 or 1 in our single occupancy subspace.  
In view of the above identity, in a translationally invariant 
ground state, what is maximized consistent with the lattice structure, 
is the valence bond amplitude. {\em Thus in the nearest neighbor 
Heisenberg models minimization of the ground state energy 
is synonymous with maximization
of the strength of nearest neighbor singlet bonds}. This proves our 
VBAM hypothesis for the spin half Mott insulator.  

Even though diagonal in terms of the number operators 
$b^\dagger_{ij} b^{}_{ij}$, equation (2) is not really diagonalized, 
as the number operators themselves do not commute whenever one
of the sites coincide:
\be
[b^\dagger_{ij} b^{}_{ij},b^\dagger_{jk} b^{}_{jk}] =
i ({\bf S}_i\times {\bf S}_k)\cdot {\bf S}_j
\ee
This non-commutativity propagates an irreducible 
minimum of bond triplet fluctuations in the ground state by 
making the ground state average $\langle b^\dagger_{ij}b^{}_{ij}
\rangle_G < 1 $.  In fact, $\langle b^\dagger_{ij}b^{}_{ij} \rangle_G 
\approx 0.6391, 0.5846 ~\mbox{and}~ = 0.5 $ respectively for 1d chain, 2d
square lattice and infinite d hyper cubic lattice. In the 1d chain
the bond triplet fluctuation is finite and manages to produce an
algebraic AFM order. In 2d it increases further and is believed to 
produce a true long range order and in infinite d it increase even
further to produce a perfect Neel order. 
The chiral operator appearing on the right hand side of equation (5) 
also tells us that chiral fluctuations are also induced in
the ground state.  
 
A long range magnetic order, when it occurs in the ground state, is 
an inevitable consequence of VBAM in the presence of the constraints 
provided by the lattice structure and the above commutation 
relation. Thus in a hyper cubic lattice for $d \geq 2$
\bearr
\mbox{VBAM + geometrical constraints} \Rightarrow 
\mbox{ AFM Order} \nonumber
\eearr
The nature of magnetic order is strongly lattice dependent. 
For the non bipartite 2d triangular lattice, 
what maximizes the bond singlet amplitude is a $120^o$ structure with 
zero point fluctuations. The case of P doped Si in the insulating state 
is described
by a 3d random lattice Heisenberg model. The nature of lattice 
constraint being very different long range magnetic order is 
believed to be absent resulting in a kind of singlet bond glass state.

The role of the long range AFM order should not be also 
overemphasized in this context.  As shown by Liang, Docout and
Anderson\cite{liang}, 
the energy difference between a disordered spin liquid
state with only short range AFM correlations and the the best 
variational state with long range AFM order is as small 
as one or two percent of the total energy:
\be
 \frac{\langle b^\dagger_{ij}b^{}_{ij}\rangle_{SL} -
 \langle b^\dagger_{ij}b^{}_{ij}\rangle_{G}} 
 {\langle b^\dagger_{ij}b^{}_{ij}\rangle_{G}}
\sim 0.01
\ee
At the level of variational wave function a small change in the 
long distance behavior of the pair function $a_{ij}$ takes us between
a disordered and ordered ground state. Hsu\cite{hsu} also shows that
the AFM order in 2d is a spinon density wave in a robust spin liquid
state.  {\em Thus the development of long range AFM order in 2d is a result 
of a small final adjustment of the VB amplitude in a spin liquid state
in the maximization procedure}. 

We wish to argue that VBAM principle continues to be approximately
valid for the optimally doped ($\delta \approx 0.15$) regime and 
produce, in the presence of the constraints modified by the hole
dynamics, the robust superconducting order and also the enhanced
charge and spin stripe correlations. The discussion 
will be heuristic and also uses known results from the RVB theory.
We will not consider the under doped regime, as disorder and long
range coulomb interaction play important roles and suppress 
superconductivity strongly and produce mesoscopic phase separation
complication. 

The t-J model
\bearr
H_{tJ} = -t \sum_{\langle ij \rangle} (c^\dagger_{i\sigma}c^{}_{j\sigma} + h.c.)
- J \sum_{\langle ij \rangle} ({\bf S}_i \cdot {\bf S}_j - 
{1\over4} n_i n_j ), \nonumber
\eearr
takes the form
\be
H_{tJ} = -t \sum_{\langle ij \rangle} (c^\dagger_{i\sigma}c^{}_{j\sigma} 
+ h.c.) - J \sum_{\langle ij \rangle } b^\dagger_{ij} b^{}_{ij} 
\ee
with the usual constraint $n_{i\uparrow} + n_{i\downarrow} \neq 2 $.
We would like to see if VBAM continues to be valid in the presence of 
the single electron hopping term in the optimally doped case.

From energy minimization
point of view the double occupancy constraint limits the single 
particle kinetic energy gain per site to $\Delta_{KE} \sim tz\delta 
$ compared to the larger super exchange energy gain $\Delta_{SE}\sim 
Jz(1-\delta^2)$. For cuprates, with a co-ordination 
number $z = 4$ and $\frac{t}{J} \approx 2$, when $\delta \sim 0.15$ 
the ratio $\frac{\Delta_{SE} }{\Delta_{KE}} \approx 3$. 
From the energy considerations given above, the VBAM principle 
continues to be important in the optimally doped regime. The above 
rough estimate is in agreement with more accurate estimates using 
variational monte carlo studies. 

The nature of constraints on valence bond
proliferation has changed now; we will see how it can stabilize 
new low temperature phases such as the robust superconductivity 
along with quasi static charge and spin stripe correlations.  
Maximizing valence bond amplitude is an important step towards
establishing a robust superconducting state.  The next important 
step is the development of `in phase resonance' of the valence bonds 
(or the zero momentum condensation of the charged valence bonds) 
in the ground state. This is what is precisely achieved in the RVB 
mean field theory. In this sense the VBAM principle is satisfied
by the RVB mean field theory.  While the original RVB mean field 
theory emphasized the extended-s mean field solution, Kotliar's 
identification of $d_{x^2-y^2}$-wave RVB solution as a 
lower energy mean field solution made the RVB theory closer 
to experiments in terms of the symmetry of the order parameter.

After the original RVB mean field theory, and some experimental
developments of that time, the inclusion of the on site
constraint was suspected to produce a large phase fluctuations
and in the even remove the finite temperature K-T transition in
an isolated $CuO_2$ layer.  
However, recent experiments\cite{kam} have strongly supported 
the one layer d-wave superconductivity with a large \tc $\approx 95 K$;
in addition the RVB Ginzburg Landau functional derived by Anderson
and the present author\cite{gaugegb,mullerhartman}, in the RVB 
gauge theory approach did not
show any singular effect on the GL coefficients due to the on site
constraints.  Very recently Lee\cite{dhlee}, 
using RVB gauge theory has provided
a rather convincing non perturbative analysis supporting the 
{\em local} stability of the d-wave solution.

What the experiments and also various theoretical clues have been
telling us is that the effect of the strong correlation (on site 
constraint) does affect superconductivity in unexpected fashions
by encouraging quasi static charge stripe and spin stripe order.
Within the VBAM principle we can understand it in the following 
heuristic fashion: the super exchange term tries to segregate the
holes so that they have more fluctuating Mott insulating region where
super exchange gain (or equivalently the bond singlet amplitude)
can be maximized.  If the hole segregation has the form of a 
1d charge stripes, in addition to VBAM, kinetic energy gain also
results from coherent charge delocalization along the charged stripes
for the following reason. In view of the Brinkman-Rice phenomenon 
and proliferation of non self retracing paths in 2d, `holes' can not 
maximize the kinetic energy gain by coherent delocalization in 2d.  
The quasi 1d fluctuating stripes on the other hand provides maximum 
of self retracing paths (along the stripes) for the holes thereby 
gaining coherent charge delocalization energy. 

Since the quasi static charge stripe formation also creates coherently
fluctuating Mott insulator region, VBAM leads to enhanced quasi 
static antiferromagnetic correlations (incommensurate order). 
Thus charge stripe and spin stripes are intimately related.
In the optimally doped region, the development of charge 
and spin stripe correlations arise from finer adjustments
of a robust superconducting state in order to satisfy the VBAM
principle. These adjustments
can be already done at the level of RVB 
mean field theory in the following fashion both in the {\em normal
state and superconducting state}: a) development of spontaneous 
anisotropic valence bond amplitudes\cite{mullerhartman}, 
$|\Delta_{x}| \neq |\Delta_{y}|$  
and b) one dimensional spatial modulation of $|\Delta_{ij}|$ either 
along the x or y axis\cite{subirstripe,dhleestripe}. 

The first case corresponds to stripes having a orientational order
(along the a or b axis) and no spatial order. In the normal state 
this will be a nematic metal with $\rho_a \neq \rho_b$ and other wise 
a non fermi liquid state. This state will have an enhanced 
magnetic correlations around $(\pi,\pi)$, in view of the spin 
localization caused by the quasi static stripe formation,
This anisotropy can continue into 
the low temperature superconducting phase and will result in a 
real mixing of $d_{x^2-y^2}$ and extended-S state. Perhaps this
phase is already seen\cite{imai} in the under doped LSCO - we think that
the conducting charge stripe glass state is an anisotropic 
metallic state (nematic metal). As explained elsewhere\cite{gbstripe}
 coupling
to octahedral rotations and displacements in LSCO further enhance
the charge stripe correlation at the expense of superconductivity. 

The second case corresponds to a spatial ordering of the stripes,
which can also start in the normal state, independent of the low
temperature superconducting state. Close to the magic filling of
$\frac{1}{8}$ this is believed to happen in doped LCO. 

In both cases there will be domain formation and the transition 
from the isotropic metallic phase to the above anisotropic phase
can be a second order phase transition.  
In the optimally doped region, even in the most favorable case of 
LSCO,  the stripe correlations are not stabilized into a true long 
range order. However, they remain as additional correlations in the
ground state there by reducing superconductivity. 

At the level of wave function modifications the above secondary
stripe orders appear as modifications of the pair function $a_{ij}$
without changing the RVB superconducting wave function in a fundamental
fashion.

We will discuss briefly the question of low energy spin fluctuations
in optimally doped cuprates.
It is an experimental fact that in the singlet dominated cuprates
there are strong low energy spin fluctuation phenomenon in k-space
particularly around the $(\pi,\pi)$ region. Among them a 
remarkably neat and unique phenomenon occurs in the bilayer cuprates: 
in neutron scattering a resolution limited 41 meV peak\cite{keimer}
 corresponding 
to a triplet excitation with momentum centered around $(\pi,\pi)$ 
is seen below the superconducting \tc.  
These excitations have very little dispersion over a wide 
momentum interval around $(\pi,\pi)$, suggesting that we can 
create spin triplet wave packets of size comparable to the lattice 
parameter as nearly exact eigen excitations. That is, the triplet 
excitation is predominantly made of nearest neighbor (Eder\cite{eder} 
also has a nearest neighbor triplet bond picture for the triplet 
resonance) triplet bonds. Availability of large valence bond amplitude 
is a prerequisite for being able to create this excited state.
It tells us that large amplitude valence  
bond exist in the ground and they can be indeed converted 
into triplets carrying a momentum of $(\pi,\pi)$ and energy 
of about 41 meV. And 41 meV is the stabilization of bond singlet
energy in the superconducting state relative to the normal state. Indeed 
various authors\cite{scalapino41,zhang41} have extracted the superconducting 
condensation energy from the spectral properties of the 
41 meV peak. It should be also mentioned that interlayer pair
tunneling adds further stability to the valence bonds in bilayer
cuprates there by supporting our VBAM principle.

We suggest that the triplet resonance around $(\pi,\pi)$ is what 
carries the memory of the Mott insulator.  It  continues to persist 
as we go to lower dopings in YBCO and the 
energy of this peak decreases linearly with the corresponding 
superconducting transition temperature.  In a Mott insulator, 
where the superconducting \tc vanishes, this triplet excitation 
around $(\pi,\pi)$ becomes soft (in view of the existing triplet
condensate) and becomes a Goldstone mode of 
the antiferromagnetic order.  This also tells us that the 
maximization of the valence bond  amplitude is easier in a doped
Mott insulator than in an undoped one. The hole dynamics in 
some sense decreases the constraints provided by the lattice,
by effectively converting the valence bond operators into 
more of `unconstrained' bosons.

Thus the various antiferromagnetic fluctuations including the 
narrow resonance peak tells us about the growing antiferromagnetic
correlation in the ground state, arising from the stabilization of
the bond singlets. It is in this sense the antiferromagnetic 
fluctuations are effects of a deep and growing bond singlet tendency, 
forced by some geometrical and dynamical constraints, rather than 
some thing that provide cooper pairing at low energies\cite{pines}.
{\em Spin fluctuations are effects rather than causes of the singlet
dominated superconductivity phenomenon}.

Thus the maximization of the bond singlet amplitude can be
thought of as a driving force behind the various anomalous
correlations one sees in cuprates including the robust high Tc 
superconductivity. The emergent quasi static charge and spin
stripe correlations and superconductivity mutually adjust 
themselves when some parameters such as doping or temperature
are changed; they are not to be thought of driving each other.  
The fundamental driving force is VBAM which is implicit in the
RVB theory.

\end{multicols}
\end{document}